# Effective field analysis using the full angular spin-orbit torque magnetometry dependence

Tomek Schulz[1,†], Kyujoon Lee[1,†], Benjamin Krüger[1], Roberto Lo Conte[1,2], Gurucharan V. Karnad[1],

Karin Garcia[3], Laurent Vila[4], Berthold Ocker[5], Dafiné Ravelosona[3], and Mathias Kläui[1,2*].

[1] Institut für Physik, Johannes Gutenberg-Universität Mainz, Staudinger Weg 7, 55128 Mainz, Germany

[2] Graduate School of Excellence "Materials Science in Mainz" (MAINZ), Staudinger Weg 9, 55128 Mainz, Germany

[3] Institut d'Electronique Fondamentale, UMR CNRS 8622, Université Paris Sud, 91405 Orsay Cedex, France

[4] Spintec, Université Grenoble Alpes, CEA-INAC, CNRS, 38054 Grenoble, France

[5] Singulus Technologies AG, 63796 Kahl am Main, Germany

[†] These authors contributed equally to this work.

[*] Correspondance and requests for materials should be addressed to M. K. (email:Klaeui@uni-mainz.de)

## Abstract

Spin-orbit torques promise ultra-efficient magnetization switching used for advanced devices based on emergent quasi-particles such as domain walls and skyrmions. Recently, the spin structure dynamics, materials and systems with tailored spin-orbit torques are being developed. A method, which allows one to detect the acting torques in a given system as a function of the magnetization direction is the torque-magnetometry method based on a higher harmonics analysis of the anomalous Hall-effect. Here we show that the effective fields acting on magnetic domain walls that govern the efficiency of their dynamics require a sophisticated analysis taking into account the full angular dependence of the torques. Using a 1-D model we compared the spin orbit torque efficiencies by depinning measurements and spin torque magnetometry. We show that the effective fields can be accurately determined and we find good agreement. Thus our method allows us now to rapidly screen materials and predict the resulting quasi-particle dynamics.

For spintronic applications, spin-orbit derived phenomena at heavy metal \ ferromagnet \ oxide (HM\FM\Ox) interfaces have become the focus of intense research efforts. In such systems the large spin-orbit coupling (SOC) in the HM layer leads to large spin-orbit torques (SOTs) acting on the magnetization in the FM in the vicinity of the HM/FM interface. This is of particular interest for promising memory technologies based on magnetic domain wall (DW) motion including the racetrack memory device [1] and DW-magnetic random access memory (DW-MRAM) devices [2]. In these concepts current-induced DW motion was originally proposed to be driven by conventional spin-transfer torque (STT), but given the much larger SOTs, these might yield even more competitive devices [3].

A key task to understand the torques e.g. the field-like torques and the damping-like torques, is to reliably determine their strengths according to their symmetries. So far, a large number of contradicting reports for the torque strengths have been put forward for nominally identical stacks. For instance, in [4] a stronger field-like than damping-like torque for a Ta\FM layer was found, whereas in [5] a negligible field-like torque was found for the same stack calling for a robust technique to measure the torques reliably.

So far, various experimental techniques have been put forward to measure the SOTs in HM\FM\Ox multilayer systems. For example, effective fields have been measured using a field-current equivalence [6–8] or higher harmonics measurements [10-15]. However, the measured values for effective fields vary depending on the technique and analysis method used. Furthermore, many analysis approaches rely on approximations, which limits their range of applicability and are often not considered in full detail making the reliability of the resulting values tenuous. Given the different reported values from different analysis methods, there is a clear need for a comprehensive comparison of these different approaches for an identical sample stack to gauge their reliability and develop a "gold standard". Additionally, the explanatory power of the obtained effective SOT- fields in a macro-spin picture for the actual effect on non-uniform magnetic textures, *i.e.,* DWs and skyrmions, needs to be clarified to properly predict the dynamics of such spin structures, as it has been predicted theoretically [16, 17] and found experimentally [14, 15] that the SOTs exhibit a strong angular dependence.

In this paper, we measure the SOT effective fields by a sophisticated higher harmonics analysis of the Hall signal taking into account the full angular dependence. Using an analytical model, we implement the angular dependence of the SOTs that allows us to rescale the SOT effective fields from the torque magnetometry measurements to describe the effective fields measured from the field-current equivalence deduced from depinning experiments. We find good agreement only when taking into account a rescaling factor.

The Si \ SiOx \ Ta (5 nm) \ $Co_{20}Fe_{60}B_{20}$ (1 nm) \ MgO (2 nm) \ Ta (5 nm) thin film was prepared by DC-and RF- sputtering using a Singulus TIMARIS/ROTARIS sputtering system. The sample was further annealed for 2 h at 300 °C in vacuum to improve the perpendicular magnetic anisotropy (PMA). The e-beam lithography process was used to pattern the nanowires. The sample was patterned into a 400 nm wide nanowire with a Hall bar and a gold Oersted line on top of the nanowire as shown in Fig. 1 (a). The depinning and spin torque magnetometry measurements were done using a 3-D cryostat where a magnetic field could be applied in the $x$, $y$ and $z$ direction. The maximum fields for the $x$ and $y$ magnets are 1 T and the maximum field for the $z$ magnet is 5 T. For the depinning measurements, the current pulses were used for generating and depinning DWs and the Hall voltage was measured by a lock-in amplifier. The spin torque magnetometry measurements were done using two lock-in amplifiers for measuring the 1$^{st}$ and 2$^{nd}$ harmonics signals.

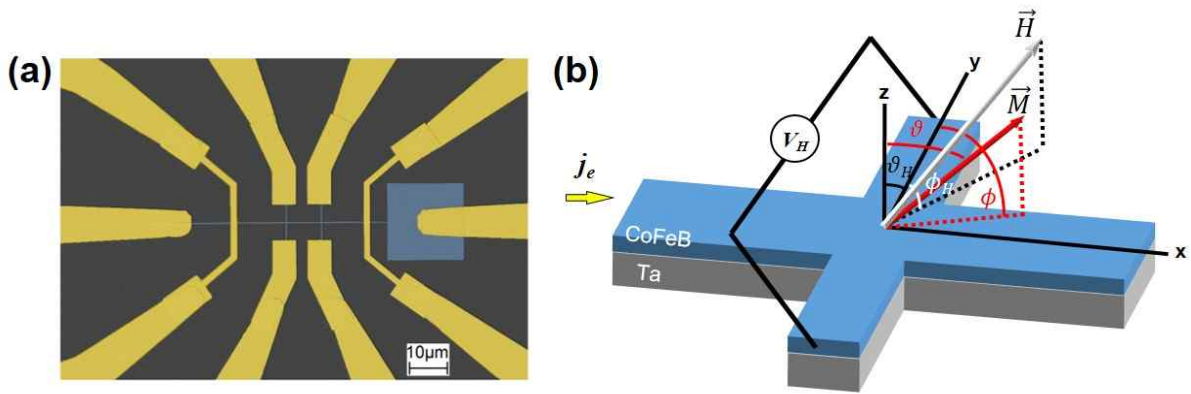

**Figure 1. (a) SEM image of the sample with the magnetic wire with two Hall crosses, a nucleation pad (highlighted in blue) and 2 Oersted field lines as well as contacts made of Au (highlighted in yellow). (b) Schematic illustration of the experimental setup and definition of the axes and angles.**

In order to calculate the effective SOT fields acting on a complex spin structure such as a domain wall, one has to take into account the angular variation of the magnetization across the spin structure. Analytically the current equivalent effective depinning field of a DW can be calculated by using the 1D model, which describes the DW in terms of two collective coordinates: the position of the DW X(t) and the azimuthal tilting angle of the DW ϕ(t). The 1D model including the adiabatic and non-adiabatic STT, the damping-like and field-like SOT, and the DMI is given by [18, 19]

$$\frac{dX}{dt} = \frac{\lambda}{(1+\alpha^2)} \left( \frac{p\pi\gamma}{2} \left( -(H_x + pH_{DMI})\sin(\phi) + (H_y + H_{FL} j)\cos(\phi) \right) \right.$$

$$\left. - p\alpha\gamma \left( H_z + \frac{\pi}{2} H_{DL} j \cos(\phi) + p\, H_{pin} + \frac{p\xi\, b_j\, j}{\gamma\lambda} \right) - \frac{1}{\lambda} b_j j + p\frac{H_k}{2}\gamma \sin(2\phi) \right)$$

and

$$\frac{d\phi}{dt} = \frac{1}{(1+\alpha^2)} \left( \frac{\pi\alpha\gamma}{2} \left( -(H_x + pH_{DMI})\sin(\phi) + (H_y + H_{FL} j)\cos(\phi) \right) \right.$$

$$\left. + \gamma \left( H_z + \frac{\pi}{2} H_{SL}\, j \cos(\phi) + p\, H_{pin} + \frac{p\xi\, b_j\, j}{\gamma\lambda} \right) - \frac{p\alpha}{\lambda} b_j j + \frac{H_k}{2}\alpha\gamma \sin(2\phi) \right).$$

Here α is the Gilbert damping parameter, γ is the gyromagnetic ratio, $H_{DL}$ and $H_{FL}$ denote the strength of the damping-like and field-like torque fields, respectively. We denote the current density with $j$, $\lambda = \sqrt{A/K_{eff}}$ is the DW width where A is the exchange stiffness constant and $K_{eff}$ is the effective anisotropy constant, $b_j = \mu_B P/(eM_s)$ is the coupling constant between the current and the magnetization, whereby $\mu_B$ is the Bohr magneton, and $e$ is the elementary charge, $P$ is the polarization and $M_s$ is the saturation magnetization. $\xi = \tau_{ex}/\tau_{sf}$, with $\tau_{ex}$ and $\tau_{sf}$ as the relaxation time of the itinerant spins due to *s-d* interaction and spin flips, respectively, $H_k$ is the DW anisotropy field, $(H_x, H_y, H_z)$ is the applied external magnetic field, $H_{DMI}$ is the effective Dzyaloshinskii-Moriya interaction field, and $H_{pin}$ is the depinning field. The parameter $p$ is +1 and -1 for a down-up and an up-down wall, respectively.

If the DW depins slowly, i.e., during a long rise time of the current pulse, the critical field $H_z$ required for depinning the DW depends on the maximum pinning field that acts upon the DW from the pinning potential. This field acts on the DW when the wall is at the position with the maximum force. If the DW overcomes this position, it is able to leave the pinning potential.

The critical field can now be derived from the following consideration: If the critical external field, $H_{dep}$ is applied, the DW will slow down approaching the position with the maximum force, move infinitely slow at the position of the maximum force, and then accelerate subsequently leading to depinning. In the 1D model, this is equivalent to a vanishing velocity ($dX/dt = 0$) and a vanishing change of the domain-wall tilting angle ($d\phi/dt = 0$) at the position with the maximum force.

We first calculate the azimuthal tilting angle of the DW in equilibrium $\phi_{eq}$. This angle can be calculated by adding the above equations. This yields

$$0 = p\pi\left(-(H_x + pH_{DMI})\sin(\phi_{eq}) + (H_y + H_{FL}j)\cos(\phi_{eq})\right) - \frac{2}{\gamma\lambda}b_j j + pH_k \sin(2\phi_{eq})$$

It is worth noting that the equilibrium tilting angle is independent of the unknown pinning field and the external field. We now insert this result in the equation for the velocity and thus obtain

$$0 = \frac{p\pi\gamma}{2}\left(-(H_x + pH_{DMI})\sin(\phi_{eq}) + (H_y + H_{FL}j)\cos(\phi_{eq})\right) - p\alpha\gamma\left(H_z + p\,H_{pin,max} + \frac{p\xi b_j j}{\gamma\lambda}\right) - \frac{p\pi\gamma\alpha}{2}H_{DL}j\cos(\phi_{eq}) - \frac{1}{\lambda}b_j j + p\frac{H_k}{2}\gamma\sin(2\phi_{eq})$$

To eliminate the unknown maximum pinning field $H_{pin,max}$ we compare the external field $H_z$ with the critical field $H_{z,0}$ when no current is applied. This yields

$$p\alpha\gamma\left(H_z - H_{z,0}\right)$$
$$= \frac{p\pi\gamma}{2}\left(-(H_x + pH_{DMI})\left(\sin(\phi_{eq}) - \sin(\phi_{eq,0})\right)\right.$$
$$+ \left(H_y + H_{FL}j\right)\left(\cos(\phi_{eq}) - \cos(\phi_{eq,0})\right)\right) - p\alpha\gamma\frac{p\xi b_j j}{\gamma\lambda} - \frac{p\pi\gamma\alpha}{2}H_{DL}j\cos(\phi_{eq})$$
$$- \frac{1}{\lambda}b_j j + p\frac{H_k}{2}\gamma\left(\sin(2\phi_{eq}) - \sin(2\phi_{eq,0})\right)$$

where $\phi_{eq,0}$ is the equilibrium tilting angle when no current is applied.

To evaluate the action of the damping like torque $H_{DL}$ on the domain wall, we calculate the depining field without applied in-plane fields, spin-transfer torque, and field-like torque. The equilibrium angle given by

$$\cos(\phi_{eq}) = p\frac{\pi}{2}\frac{H_{DMI}}{H_k} \quad \text{or} \quad \sin(\phi_{eq}) = 0$$

then becomes independent of the current simplifying the expression

$$\Delta H_z = H_{dep} - H_{z,0} = -\frac{\pi}{2}H_{DL}j$$

(1)

for the critical external field. We thus see that the effect of the field $H_{DL}$ from the damping-like torque on the domain wall is by a factor $\pi/2$ larger than the effect of an external field.[20] This is important to note, as it vitiates the previously used expectation that $H_z$ is equal to the effective field corresponding to the SOTs. Here we show that one needs to rescale the field by a factor $\pi/2$ due to the fact that in a Néel wall $H_{DL}$ is everywhere perpendicular to the magnetization, thus leading to a maximum torque, while the external field is not. So while this calculation shows that the effective fields need to be rescaled, the fixed factor determined here relies on a torque that is constant and does not exhibit a strong angular dependence, which might not hold for all systems.

The measured Hall resistance $R_{xy}$ can be described as follows,

$$R_{xy} = \tfrac{1}{2}\Delta R_{AHE}\cos\vartheta + \tfrac{1}{2}\Delta R_{PHE}\sin^2\vartheta\sin 2\phi. \tag{2}$$

Here, $\vartheta$ and $\phi$ are the polar and azimuthal angle, respectively (see Fig. 1(b) for their definition). $R_{AHE}$ and $R_{PHE}$ are the anomalous Hall resistance and planar Hall resistance, respectively. When a sinusoidal current $I = \Delta I \sin \omega t$, with a frequency $\omega$ that is low compared to the electronic relaxation and spin dynamics time scales, is applied to sample, the corresponding current-induced effective fields oscillate with the current. In this study, we use a frequency $\omega = 13.7$ Hz, so that the phase difference between the current and the current-induced effective fields is negligible.

When an in-plane magnetic field is applied to the sample, the magnetization direction can be tilted away from the out-of-plane easy axis direction leading to a new equilibrium position defined by the polar angle $\vartheta_o$ and the azimuthal angle $\phi_o$ (see Fig. 1 (b)). Without any PHE, the measured Hall resistance is directly proportional to the z-component of the magnetization, $R_{AHE} \sim m_z = \cos \vartheta_o$. This can be used to evaluate the magnetization direction at equilibrium ($\vartheta_o, \phi_o$) set by the external applied field competing with the internal anisotropy field. The SOTs originating from the net non-equilibrium spin-density formed at the HM/FM interface and the bulk SHE exert a torque on the magnetization in equilibrium position leading to an extra tilt of the magnetization $\Delta\vartheta$ and $\Delta\phi$ away from $\vartheta_o$ and $\phi_o$.

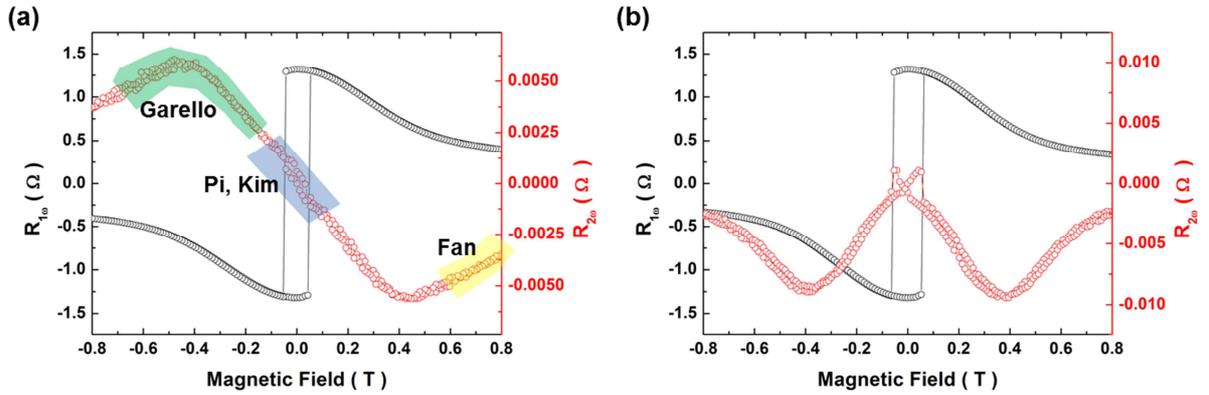

**Figure 2. Schematic illustration of the different measurement and analysis methods fitting regime. (a) and (b) show the first (black) and second harmonics (red) Hall voltage signals as a function of an applied external magnetic in-plane field in the longitudinal and transversal direction, respectively. The different color marks show the different regimes using different analysis methods. Blue indicates the Pi [10], Kim [11] method, Green the Garello [14] method, and yellow the Fan [13] method.**

In our analysis, we utilize three different analyses in order to get a full angular dependence of the SOTs [10-15]. In Fig. 2 the first (black line) and second (red line) harmonics signal of the Hall voltage are depicted for an in-plane magnetic field scan in the longitudinal ($\phi_0 = 90°$) (Fig. 2 (a)) and transversal ($\phi_0 = 0°$) (Fig. 2(b)) direction, respectively. Furthermore, we illustrate in Fig. 2 (a) schematically the field regime that we focus on for the different approaches within their validity ranges. As the application of an external in-plane magnetic field tilts the magnetization equilibrium direction away from the easy axis towards the plane, the different measurement and analysis methods can be distinguished by their different polar angle regimes of validity.

Using the first method reported [10], the SOTs are measured in the small angle regime, close to the easy axis direction of the PMA system and only for the transversal in-plane field direction. Here the first (second) harmonics signal follow a quadratic (linear) dependence on the applied external in-plane field. A small angle approximation can be used in the derivation of the analytical expression of the $1\omega$ and $2\omega$ voltages, which is the basis of methods focusing on the small angle regime. An extension of this first analysis have been reported, e.g. Kim et al. [11] who measured the SOTs not only for the transversal direction, which was initially attributed only to the ISGE (also termed Rashba-Edelstein effect) effective field, but also for the longitudinal direction, which was identified initially only as resulting from the SHE-STT. Recent theories [21, 22] consider an intermixing of both, so that the ISGE as well as the SHE can contribute to both the longitudinal and the transversal effective field acting on the magnetization.

More recently, the PHE contribution in the measurement scheme has been taken into account as an additional contribution to the Hall effect [12, 14, 23]. Using $1\omega$ and $2\omega$ voltages [12] for a PMA system, the ratio between the 1$^{st}$ derivative of the $2\omega$ signal and the 2$^{nd}$ derivative of the $1\omega$ signal with respect to the applied external field can be expressed as,

$$H_{x,y} = \left( \frac{\partial V_{2\omega}}{\partial H_{x,y}} \middle/ \frac{\partial^2 V_{\omega}}{\partial H_{x,y}^2} \right) \qquad (3)$$

The effective SOT fields per current density are then given by

$$\mu_0 h_{DL} = -\frac{2\frac{(B_X \pm 2\varepsilon B_Y)}{1-4\varepsilon^2}}{j}, \quad \mu_0 h_{FL} = -\frac{2\frac{(B_Y \pm 2\varepsilon B_X)}{1-4\varepsilon^2}}{j}, \quad (4)$$

Here, $\varepsilon = \frac{\Delta R_{PHE}}{\Delta R_{AHE}}$ reflects in which extend the change in the $1\omega$ and $2\omega$ signal comes from the PHE and AHE. In our stack $\varepsilon \sim 5\%$. Using this method we obtain $\mu_0 h_{DL} = (-1.43 \pm 0.04)$ mT/$10^{11}$ Am$^{-2}$ and $\mu_0 h_{FL} = (-1.88 \pm 0.05)$ mT/$10^{11}$ Am$^{-2}$ for the polar angular regime $0° < \vartheta < 10°$. Without the consideration of the PHE, we find $\mu_0 h_{DL} = (-1.60 \pm 0.05)$ mT/$10^{11}$ Am$^{-2}$ and $\mu_0 h_{FL} = (-2.01 \pm 0.05)$ mT/$10^{11}$ Am$^{-2}$, which gives a relative error of 12 % and 7 % for $\Delta h_{DL}$ and $\Delta h_{FL}$, respectively.

For polar angles $\vartheta \gtrsim 10°$, the first and second harmonic signal deviate from the quadratic and linear dependence, respectively, though the small angle approximation cannot be used anymore. Instead, the second harmonic Hall voltages show a non-trivial angular dependence, which reflects the anisotropy of the SOTs. These anisotropies have been investigated experimentally first by Garello et al. [14] for the angular regime $10° < \vartheta < 60°$. Instead of using the approximation of a quadratic angular dependence of the equilibrium angle $\vartheta_0$ on the external field, the complex dependence of $\vartheta_0(B_{ext})$ is taken into account, by using the derivative of the first harmonics signal. This leads to the following expressions for second harmonic signal:

$$V_H^{2\omega} = (\Delta R_{AHE} - 2\Delta R_{PHE} \cos\vartheta \sin 2\varphi) \frac{d\cos\vartheta}{dB_{ext}} \frac{B_\vartheta}{\sin(\vartheta_H - \vartheta)} + 2\Delta R_{PHE} \sin^2\vartheta \cos 2\varphi \frac{B_\varphi}{\sin(\vartheta_H)}, \quad (5)$$

By fitting the experimental data into Eq. (7), we derive the effective SOT fields $\mu_0 h_{DL}$ and $\mu_0 h_{FL}$, as depicted in Fig. 3(a).

For large in-plane magnetic fields applied to the sample the equilibrium direction of the magnetization vector is pointing almost into the plane and the second harmonics signal can be approximated as described in [13]:

$$R_H^{2\omega} = -\frac{1}{2} \frac{\Delta R_{AHE} H_{DL,FL}}{(H_{ext} - H_K)}. \quad (6)$$

Where $H_K$ is the anisotropy field. With this approach we derive $\mu_0 h_{DL} = (-1.29 \pm 0.1)$ mT/$10^{11}$ Am$^{-2}$ and $\mu_0 h_{FL} = (-0.11 \pm 0.04)$ mT/$10^{11}$ Am$^{-2}$ for the polar angular regime $\vartheta < 75°$. Since in this analysis the planar Hall effect is not taken into account. In our system the planar Hall effect is only 5% of the anomalous Hall effect, therefore we can still apply the method in calculating the effective fields in the high magnetic field regime without generating large errors.

As we employ the different analysis methods for the first time to the same sample, we can clearly see from the deviating results, that one needs to take into account the validity regime for each method.

Having analyzed all these methods relying on higher harmonics magnetometry, we can now put together the results obtained for specific angular regimes and derive a full angular dependence $H'_{DL}(\vartheta)$ and $H'_{FL}(\vartheta)$ as the combination of the individual results. All individual results (solid lines and symbols), as well as the generated combined full angular dependence (dashed lines) are shown in Fig. 3. Here, the full angular dependence is extracted as a simple polynomial fit of the individual curves and symbols. Additionally, we have calculated the effective spin Hall angle θ$_{SH}$ from the damping-like effective field H$_{DL}$. Using the relation θ$_{SH}$ = ($\mu_o$(H$_{DL}$/$j_e$)$e$M$_s t_{FM}$)/$\hbar$ [23,24] where $\mu_o$ is the magnetic permeability, $j_e$ is the current density, M$_s$ = 1.1 × 10$^6$ A/m, $t_{FM}$ is the thickness of the CoFeB, and $\hbar$ is the Planck constant, we were able to calculate the effective spin Hall angles. The calculated maximum effective spin Hall angle is 0.076 for the direction with the largest torque. This value is in the range of previously reported values [25, 26].

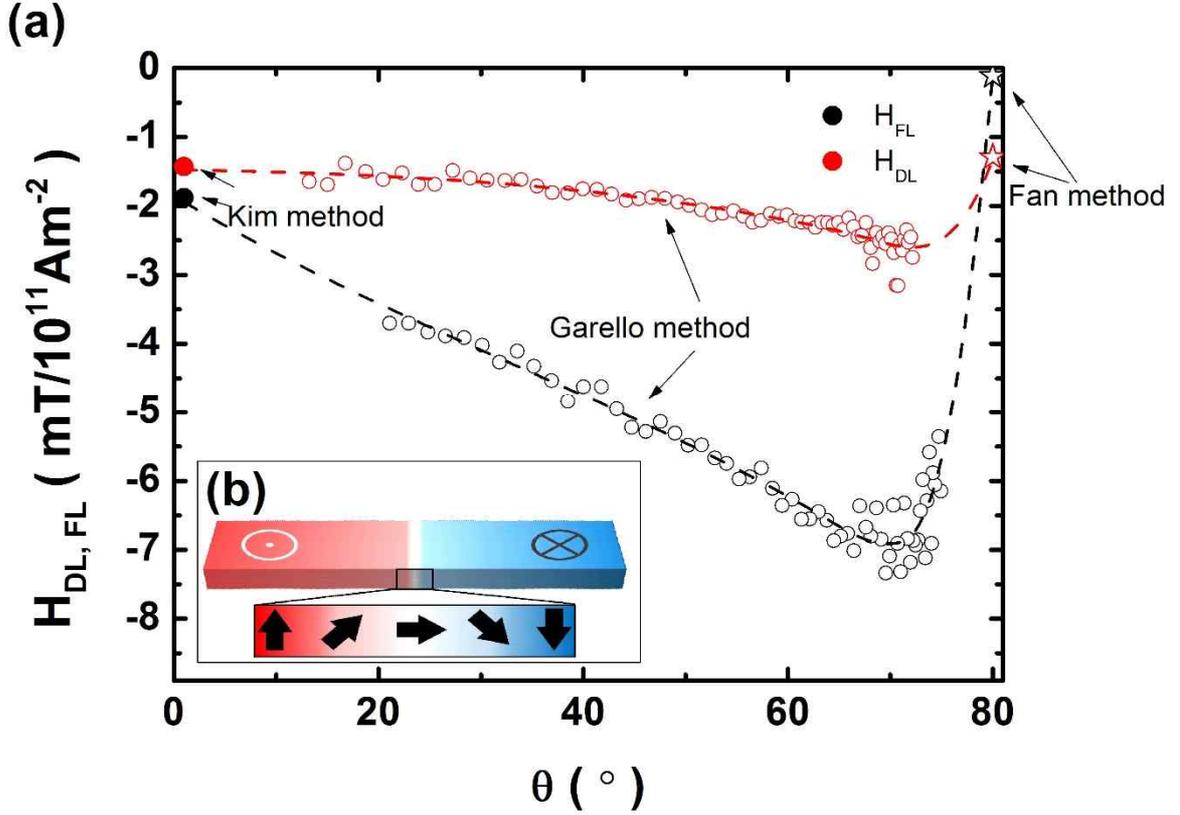

**Figure 3. (a) Angular dependence of the SOTs obtained by using three different analysis procedures for three different angular regimes of the magnetization angle Θ (red is $H_{DL}$ and black is $H_{FL}$). We see that the methods cover their respective angular validity ranges and together provide a reasonably good coverage of the full angular range. (b) Illustration of the internal magnetization direction of a Néel type DW. The polar angle changes from 0° to $\pi$ as one follows the DW profile from left to right.**

We see that the combination of the methods allows us to determine the full angular dependence. The final aim is now to use this combined full angular dependence of the SOT effective fields, to calculate the expected DW depinning fields measured from the current-field equivalence. For this we take into account the DW profile, as the internal DW magnetization direction varies along the DW. The DW profile is described by

$$\boldsymbol{M} = \frac{M_s}{\cosh\frac{x-X}{\lambda}} \begin{pmatrix} \cos\phi \\ \sin\phi \\ \sinh\frac{x-X}{\lambda} \end{pmatrix}$$

and is illustrated schematically for a Néel type DW ($\phi = 90°$) and $X = 0$ in Fig. 3(b).

Next we calculate the average effective SOT field by integrating the individual SOTs as function of DW position, using

$$H_{DL,FL}^{avg} = \frac{1}{\pi\lambda} \int \frac{H_{DL,FL}(\vartheta(x))}{\cosh\left(\frac{x-X}{\lambda}\right)} dx,$$

whereby $H_{DL,FL}(\vartheta(x))$ is the generated combined angular dependence fit. Using this approach we determine the averaged effective SOT-fields per current density $\mu_0 h_{DL}^{avg} = -1.57$ mT / $10^{11}$ Am⁻² and $\mu_0 h_{FL}^{avg} = -4.56$ mT / $10^{11}$ Am⁻².

Finally, we use these values for the effective SOT fields to calculate the current equivalent out-of-plane field per current density $\mu_0 \Delta h_{z,}^{cal} \sim (2.42 \pm 0.5)$ mT / $10^{11}$ Am⁻² which is in good agreement within the error bars with the results obtain by the depinning experiments for current applied perpendicular to the DW, yielding $\mu_0 \Delta h_z^{exp} \sim (2.34 \pm 0.63)$ mT / $10^{11}$ Am⁻² (details of the methods see [17]). It is noticeable that the current equivalent field calculated from the SOT field measured at the low angle magnetization tilt show a value of $\mu_0 \Delta h_{z,}^{cal} \sim (2.24 \pm 0.7)$ mT / $10^{11}$ Am⁻² since the angular dependence of the damping-like effective field is small. It is also noticeable that although the field like effective field has a large angular dependence the value of the field-like effective field does not have a strong effect on the final value of the converted effective fields for the depinning measurements. This is due to the domain wall depinning not being very susceptible to the field-like torque and in the simplest rigid wall model, the depinning only depends on the damping-like torque as shown in equation (1). This shows that using the effective fields determined from the torque magnetometry measurements can be used as a reliable gauge of the effective DW depinning fields if one takes into account the conversion factor between the measured change in the field and the SOT effective field, as well as the complex angular dependence of the SOTs itself. Our analysis thus paves the way to predict the equivalent fields acting on a spin structure and thus predict the spin dynamics, which is a key goal for the understanding of the current-induced dynamics of magnetic quasi-particles, such as domain walls and skyrmions in these systems.

In summary, we have developed the necessary sophisticated analysis method to determine the effects of SOT fields determined by higher harmonic analysis acting on emergent quasi particle spin structures such as DWs or skyrmions. Using a collective coordinate's model, we find that the fields determined by the higher harmonics analysis need to be rescaled to compare them with effective fields determined from the current-field equivalence of depinning measurements. We analyze the angular dependence of the SOTs by combining three different analysis methods. From this we determine experimentally the full angular dependence of the SOTs with the polar angle. Finally, the full angular dependence of the torques is used to integrate out the total effective field acting on a domain wall during a depinning experiment. This is necessary since in DWs we have spin texture with a full angular rotation. We compare the value deduced from our analysis method to the value measured experimentally in a depinning experiment on the same sample and find good agreement only when using the new method showing the relevance of our approach. The developed method now opens a path to using the relatively simple SOT magnetometry based on high harmonics analysis for the first time to reliably predict the effective fields acting on spin structures such as DWs or skyrmions. As these are the relevant emergent quasi-particles proposed for many applications, we can now use our method to rapidly screen materials and predict the quasi-particle dynamics using simple collective coordinates models as well as compare to depinning field measurements to reconcile measurements using the different approaches.

We like to thank K. Garello for fruitful discussions. We acknowledge support by the Graduate School of Excellence Materials Science in Mainz (MAINZ) GSC 266, Staudinger Weg 9, 55128, Germany; the DFG (KL1811, SFB TRR 173 Spin+X); the EU (MultiRev ERC-2014-PoC 665672; WALL, FP7-PEOPLE-2013-ITN 608031) and the Research Center of Innovative and Emerging Materials at Johannes Gutenberg University (CINEMA). K. L. acknowledges the European Research Council (ERC) under the European Union's Horizon 2020 research and innovation programme (Grant Agreement No. 709151).

T.S. and K.L. contributed equally to this work.

# REFERENCES

[1] Parkin, S. S. P., Hayashi, M. & Thomas, L. Magnetic domain-wall racetrack memory. *Science* **320,** 190–194 (2008).

[2] Fukami, S. *et al.* Low-Current Perpendicular Domain Wall Motion Cell for Scalable High-Speed MRAM. *2009 Symposium on VLSI Technology Digest of Technical Papers* 230–231 (2009).

[3] Parkin, S. & Yang, S.-H. Memory on the racetrack. *Nature Nanotechnology* **10,** 195–198 (2015).

[4] Torrejon, J. *et al.* Interface control of the magnetic chirality in CoFeB/MgO heterostructures with heavy-metal underlayers. *Nature Communications* **5,** (2014).

[5] Liu, L. *et al.* Spin-torque switching with the giant spin Hall effect of tantalum. *Science* **336,** 555–558 (2012).

[6] L. Liu, O. J. Lee, T. J. Gudmundsen, D. C. Ralph, and R. A. Buhrman, Phys. Rev. Lett. 109, 096602 (2012)

[7] Heinen, J. *et al.* Current-induced domain wall motion in Co/Pt nanowires: Separating spin torque and Oersted-field effects. *Applied Physics Letters* **96,** 202510 (2010).

[8] Schulz, T. *et al.* Spin-orbit torques for current parallel and perpendicular to a domain wall. *Applied Physics Letters* **107,** 122405 (2015).

[9] Pi, U. H. *et al.* Tilting of the spin orientation induced by Rashba effect in ferromagnetic metal layer. *Appl. Phys. Lett.* **97,** 162507 (2010).

[10] Kim, J. *et al.* Layer thickness dependence of the current-induced effective field vector in Ta|CoFeB|MgO. *Nat Mater* **12,** 240–245 (2013).

[11] Hayashi, M., Kim, J., Yamanouchi, M. & Ohno, H. Quantitative characterization of the spin-orbit torque using harmonic Hall voltage measurements. *Phys. Rev. B* **89,** 144425 (2014).

[12] Fan, Y. *et al.* Magnetization switching through giant spin–orbit torque in a magnetically doped topological insulator heterostructure. *Nat Mater* **13,** 699–704 (2014).

[13] Garello, K. *et al.* Symmetry and magnitude of spin-orbit torques in ferromagnetic heterostructures. *Nat Nano* **8,** 587–593 (2013).

[14] Qiu, X. *et al.* Angular and temperature dependence of current induced spin-orbit effective fields in Ta/CoFeB/MgO nanowires. *Scientific Reports* **4,** (2014).

[15] Ortiz Pauyac, C., Wang, X., Chshiev, M. & Manchon, A. Angular dependence and symmetry of Rashba spin torque in ferromagnetic heterostructures. *Applied Physics Letters* **102,** 252403 (2013).

[16] Lee, K.-S. *et al.* Angular dependence of spin-orbit spin-transfer torques. *Physical Review B* **91,** (2015).

[17] Jué, E. *et al.* Chiral damping of magnetic domain walls. *Nat. Mater*. **15**, 272 (2016)

[18] Emori, S., Bauer, U., Ahn, S.-M., Martinez, E. & Beach, G. S. D. *Nature Materials* **12,** 611–616 (2013).

[19] O. Boulle, L. D. Buda-Prejbeanu, E. Jué, I. M. Miron, & G. Gaudin, *Journal of Applied Physics* **115,** 17D502 (2014).

[20] Thiaville, A., Rohart, S., Jue, E., Cros, V. & Fert, A. *Europhysics Letters*, **100**, 57002 (2012).

[21] Kurebayashi, H. *et al. Nature Nanotechnology* **9,** 211–217 (2014).

[22] Haney, P. M., Lee, H.-W., Lee, K.-J., Manchon, A. & Stiles, M. D. *Physical Review B* **88,** (2013).

[23] Lee, H.-R. *et al. Scientific Reports* **4,** 6548 (2014).

[24] X. Qiu, P. Deorani, K. Narayanapillai, K.-S. Lee, K.-J.Lee, H.-W. Lee, and H. Yang, *Scientific Reports* **4**, 4491 (2014).

[25] L. Liu, C. Pai, Y. Li, H. W. Tseng, D. C. Ralph, and R. A. Buhrman, Science 336, 555 (2012).

[26] T. Seifert, et al. *Nature Photonics* **10**, 483 (2016).